\documentclass[aps,prd,twocolumn,superscriptaddress,showpacs,preprintnumbers,amsmath,amssymb]{revtex4}

\usepackage{graphicx}
\usepackage{dcolumn}
\usepackage{bm}
\usepackage[dvips]{color}
\usepackage{epsfig}

\begin{document}

\newcommand{\mbc}{M_{\rm bc}}
\newcommand{\de}{\Delta E}
\newcommand{\thr}{\cos\theta_{\rm thr}}
\newcommand{\fish}{\mathcal{F}}

\newcommand{\bdk}{$B^{\pm}\to DK^{\pm}$}
\newcommand{\bdkm}{$B^{-}\to DK^{-}$}
\newcommand{\bdkp}{$B^{+}\to DK^{+}$}
\newcommand{\bdtk}{$B^{\pm}\to \tilde{D}K^{\pm}$}
\newcommand{\bdtkp}{$B^{+}\to \tilde{D}_+K^{+}$}
\newcommand{\bdtkm}{$B^{-}\to \tilde{D}_-K^{-}$}

\newcommand{\bdsk}{$B^{\pm}\to D^{*}K^{\pm}$}
\newcommand{\bdskm}{$B^{-}\to D^{*}K^{-}$}
\newcommand{\bdskp}{$B^{+}\to D^{*}K^{+}$}
\newcommand{\bdstk}{$B^{\pm}\to \tilde{D}^{*}K^{\pm}$}
\newcommand{\bdstkp}{$B^{+}\to \tilde{D}^{*}_+K^{+}$}
\newcommand{\bdstkm}{$B^{-}\to \tilde{D}^{*}_-K^{-}$}

\newcommand{\bdks}{$B^{\pm}\to DK^{*\pm}$}
\newcommand{\bdksm}{$B^{-}\to DK^{*-}$}
\newcommand{\bdksp}{$B^{+}\to DK^{*+}$}
\newcommand{\bdtks}{$B^{\pm}\to \tilde{D}K^{*\pm}$}
\newcommand{\bdtksp}{$B^{+}\to \tilde{D}_+K^{*+}$}
\newcommand{\bdtksm}{$B^{-}\to \tilde{D}_-K^{*-}$}

\newcommand{\bddsk}{$B^{\pm}\to D^{(*)}K^{\pm}$}
\newcommand{\bddskm}{$B^{-}\to D^{(*)}K^{-}$}
\newcommand{\bddskp}{$B^{+}\to D^{(*)}K^{+}$}
\newcommand{\bddstk}{$B^{\pm}\to \tilde{D}^{(*)}K^{\pm}$}
\newcommand{\bddstkp}{$B^{+}\to \tilde{D}^{(*)}_+K^{+}$}
\newcommand{\bddstkm}{$B^{-}\to \tilde{D}^{(*)}_-K^{-}$}

\newcommand{\bddsks}{$B^{\pm}\to D^{(*)}K^{(*)\pm}$}
\newcommand{\bddsksm}{$B^{-}\to D^{(*)}K^{(*)-}$}
\newcommand{\bddsksp}{$B^{+}\to D^{(*)}K^{(*)+}$}
\newcommand{\bddstks}{$B^{\pm}\to \tilde{D}^{(*)}K^{(*)\pm}$}
\newcommand{\bddstksp}{$B^{+}\to \tilde{D}^{(*)}_+K^{(*)+}$}
\newcommand{\bddstksm}{$B^{-}\to \tilde{D}^{(*)}_-K^{(*)-}$}

\newcommand{\bdksnr}{$B^{\pm}\to DK^0_S\pi^{\pm}$}
\newcommand{\bdkspnr}{$B^{+}\to DK^0_S\pi^{+}$}
\newcommand{\bdksmnr}{$B^{-}\to DK^0_S\pi^{-}$}
\newcommand{\bdtksnr}{$B^{\pm}\to \tilde{D}K^0_S\pi^{\pm}$}
\newcommand{\bdtkspnr}{$B^{+}\to \tilde{D}_+K^0_S\pi^{+}$}
\newcommand{\bdtksmnr}{$B^{-}\to \tilde{D}_-K^0_S\pi^{-}$}

\newcommand{\bdpi}{$B^{\pm}\to D\pi^{\pm}$}
\newcommand{\bdpip}{$B^{+}\to D\pi^{+}$}
\newcommand{\bdpim}{$B^{-}\to D\pi^{-}$}
\newcommand{\bdtpi}{$B^{\pm}\to \tilde{D}\pi^{\pm}$}
\newcommand{\bdtpip}{$B^{+}\to \tilde{D}_+\pi^{+}$}
\newcommand{\bdtpim}{$B^{-}\to \tilde{D}_-\pi^{-}$}

\newcommand{\bdstpi}{$B^{\pm}\to \tilde{D}^{*}\pi^{\pm}$}
\newcommand{\bdstpip}{$B^{+}\to \tilde{D}^{*}_+\pi^{+}$}
\newcommand{\bdstpim}{$B^{-}\to \tilde{D}^{*}_-\pi^{-}$}
\newcommand{\bdspi}{$B^{\pm}\to D^{*}\pi^{\pm}$}
\newcommand{\bdspim}{$B^{-}\to D^{*}\pi^{-}$}
\newcommand{\bdspip}{$B^{+}\to D^{*}\pi^{+}$}

\newcommand{\bndspi}{$B\to D^{*\pm}\pi^{\mp}$}
\newcommand{\bndspim}{$B\to D^{*+}\pi^{-}$}
\newcommand{\bndspip}{$B\to D^{*-}\pi^{+}$}

\newcommand{\bddspi}{$B^{\pm}\to D^{(*)}\pi^{\pm}$}
\newcommand{\bddspip}{$B^{+}\to D^{(*)}\pi^{+}$}
\newcommand{\bddspik}{$B^{-}\to D^{(*)}\pi^{-}$}
\newcommand{\bddstpi}{$B^{\pm}\to \tilde{D}^{(*)}\pi^{\pm}$}
\newcommand{\bddstpip}{$B^{+}\to \tilde{D}^{(*)}_+\pi^{+}$}
\newcommand{\bddstpim}{$B^{-}\to \tilde{D}^{(*)}_-\pi^{-}$}

\newcommand{\dsdpi}{$D^{*\pm}\to D\pi^{\pm}$}
\newcommand{\dsdpis}{$D^{*\pm}\to D\pi_s^{\pm}$}
\newcommand{\dsdpip}{$D^{*+}\to D^0\pi^{+}$}
\newcommand{\dsdpips}{$D^{*+}\to D^0\pi_s^{+}$}
\newcommand{\dsdpim}{$D^{*-}\to \overline{D}{}^0\pi^{-}$}
\newcommand{\dsdpims}{$D^{*-}\to \overline{D}{}^0\pi_s^{-}$}

\newcommand{\dkpp}{$\overline{D}{}^0\to K^0_S\pi^+\pi^-$}
\newcommand{\dtkpp}{$\tilde{D}\to K^0_S\pi^+\pi^-$}

\newcommand{\new}[1]{\textcolor{red}{#1}}
\newcommand{\bmu}{\boldsymbol{\mu}}
\newcommand{\bz}{\boldsymbol{z}}

\renewcommand{\arraystretch}{1.4}

\preprint{BELLE-CONF-0801}

\title{Updated Measurement of \boldmath{$\phi_3$} with a Dalitz Plot
Analysis of \boldmath{\bddskp} Decay}

\affiliation{Budker Institute of Nuclear Physics, Novosibirsk}
\affiliation{Chiba University, Chiba}
\affiliation{University of Cincinnati, Cincinnati, Ohio 45221}
\affiliation{Department of Physics, Fu Jen Catholic University, Taipei}
\affiliation{Justus-Liebig-Universit\"at Gie\ss{}en, Gie\ss{}en}
\affiliation{The Graduate University for Advanced Studies, Hayama}
\affiliation{Gyeongsang National University, Chinju}
\affiliation{Hanyang University, Seoul}
\affiliation{University of Hawaii, Honolulu, Hawaii 96822}
\affiliation{High Energy Accelerator Research Organization (KEK), Tsukuba}
\affiliation{Hiroshima Institute of Technology, Hiroshima}
\affiliation{University of Illinois at Urbana-Champaign, Urbana, Illinois 61801}
\affiliation{Institute of High Energy Physics, Chinese Academy of Sciences, Beijing}
\affiliation{Institute of High Energy Physics, Vienna}
\affiliation{Institute of High Energy Physics, Protvino}
\affiliation{Institute for Theoretical and Experimental Physics, Moscow}
\affiliation{J. Stefan Institute, Ljubljana}
\affiliation{Kanagawa University, Yokohama}
\affiliation{Korea University, Seoul}
\affiliation{Kyoto University, Kyoto}
\affiliation{Kyungpook National University, Taegu}
\affiliation{\'Ecole Polytechnique F\'ed\'erale de Lausanne (EPFL), Lausanne}
\affiliation{University of Ljubljana, Ljubljana}
\affiliation{University of Maribor, Maribor}
\affiliation{University of Melbourne, School of Physics, Victoria 3010}
\affiliation{Nagoya University, Nagoya}
\affiliation{Nara Women's University, Nara}
\affiliation{National Central University, Chung-li}
\affiliation{National United University, Miao Li}
\affiliation{Department of Physics, National Taiwan University, Taipei}
\affiliation{H. Niewodniczanski Institute of Nuclear Physics, Krakow}
\affiliation{Nippon Dental University, Niigata}
\affiliation{Niigata University, Niigata}
\affiliation{University of Nova Gorica, Nova Gorica}
\affiliation{Osaka City University, Osaka}
\affiliation{Osaka University, Osaka}
\affiliation{Panjab University, Chandigarh}
\affiliation{Peking University, Beijing}
\affiliation{University of Pittsburgh, Pittsburgh, Pennsylvania 15260}
\affiliation{Princeton University, Princeton, New Jersey 08544}
\affiliation{RIKEN BNL Research Center, Upton, New York 11973}
\affiliation{Saga University, Saga}
\affiliation{University of Science and Technology of China, Hefei}
\affiliation{Seoul National University, Seoul}
\affiliation{Shinshu University, Nagano}
\affiliation{Sungkyunkwan University, Suwon}
\affiliation{University of Sydney, Sydney, New South Wales}
\affiliation{Tata Institute of Fundamental Research, Mumbai}
\affiliation{Toho University, Funabashi}
\affiliation{Tohoku Gakuin University, Tagajo}
\affiliation{Tohoku University, Sendai}
\affiliation{Department of Physics, University of Tokyo, Tokyo}
\affiliation{Tokyo Institute of Technology, Tokyo}
\affiliation{Tokyo Metropolitan University, Tokyo}
\affiliation{Tokyo University of Agriculture and Technology, Tokyo}
\affiliation{Toyama National College of Maritime Technology, Toyama}
\affiliation{Virginia Polytechnic Institute and State University, Blacksburg, Virginia 24061}
\affiliation{Yonsei University, Seoul}
  \author{K.~Abe}\affiliation{High Energy Accelerator Research Organization (KEK), Tsukuba} 
  \author{I.~Adachi}\affiliation{High Energy Accelerator Research Organization (KEK), Tsukuba} 
  \author{H.~Aihara}\affiliation{Department of Physics, University of Tokyo, Tokyo} 
  \author{K.~Arinstein}\affiliation{Budker Institute of Nuclear Physics, Novosibirsk} 
  \author{T.~Aso}\affiliation{Toyama National College of Maritime Technology, Toyama} 
  \author{V.~Aulchenko}\affiliation{Budker Institute of Nuclear Physics, Novosibirsk} 
  \author{T.~Aushev}\affiliation{\'Ecole Polytechnique F\'ed\'erale de Lausanne (EPFL), Lausanne}\affiliation{Institute for Theoretical and Experimental Physics, Moscow} 
  \author{T.~Aziz}\affiliation{Tata Institute of Fundamental Research, Mumbai} 
  \author{S.~Bahinipati}\affiliation{University of Cincinnati, Cincinnati, Ohio 45221} 
  \author{A.~M.~Bakich}\affiliation{University of Sydney, Sydney, New South Wales} 
% ...
\collaboration{The Belle Collaboration}
\noaffiliation

\begin{abstract} 

We present an updated measurement of the unitarity triangle angle $\phi_3$ 
using a 
Dalitz plot analysis of the $K^0_S\pi^+\pi^-$ decay of the neutral $D$ meson 
produced in \bddsk\ decays. The method exploits the interference between 
$D^0$ and $\overline{D}{}^0$ to extract the angle $\phi_3$, strong phase 
$\delta$ and the ratio $r$ of suppressed and allowed amplitudes. We apply 
this method to a 605 fb$^{-1}$ data sample collected by the Belle experiment. 
The analysis uses two modes: \bdkp, and \bdskp\ with $D^{*}\to D\pi^0$, 
as well as the corresponding charge-conjugate modes. 
From a combined maximum likelihood fit to the two modes, we obtain 
$\phi_3=76^{\circ}\;^{+12^{\circ}}_{-13^{\circ}}
\mbox{(stat)}\pm 4^{\circ} \mbox{(syst)}\pm 9^{\circ}(\mbox{model})$. 
The statistical significance of $CP$ violation ($\phi_3\neq 0$) in 
our measurement is $(1-5.5\times 10^{-4})$, or 3.5 standard deviations. 
These results are preliminary. 
\end{abstract}
\maketitle

\tighten

\section{Introduction}

Determinations of the Cabibbo-Kobayashi-Maskawa
(CKM) \cite{ckm} matrix elements provide important checks on
the consistency of the standard model and ways to search
for new physics. The possibility of observing direct $CP$ violation
in $B\to D K$ decays was first discussed by I. Bigi, A. Carter and 
A. Sanda \cite{bigi}.
Since then, various methods using $CP$ violation in $B\to D K$ decays have been
proposed \cite{glw,dunietz,eilam,ads} to measure the unitarity triangle
angle $\phi_3$. Three body final states such as 
$K^0_S\pi^+\pi^-$ \cite{giri,binp_dalitz} have been suggested as 
promising modes for the extraction of $\phi_3$. 
In the Wolfenstein parameterization of the CKM matrix elements, 
the weak parts of the amplitudes 
that contribute to the decay \bdkp\ 
are given by $V_{cb}^*V_{us\vphantom{b}}^{\vphantom{*}}\sim A\lambda^3$ 
(for the $\overline{D}{}^0 K^+$ final state) and
$V_{ub}^*V_{cs\vphantom{b}}^{\vphantom{*}}\sim A\lambda^3(\rho+i\eta)$ (for $D^0 K^+$). 
The two amplitudes interfere as the $D^0$ and $\overline{D}{}^0$ mesons decay
into the same final state $K^0_S \pi^+ \pi^-$. 
Assuming no $CP$ asymmetry in neutral $D$ decays, 
the amplitude of the neutral $D$ decay from \bdk\ 
as a function of Dalitz plot variables $m^2_+=m^2_{K^0_S\pi^+}$ and 
$m^2_-=m^2_{K^0_S\pi^-}$ is 
\begin{equation}
  M_{\pm}=f(m^2_{\pm}, m^2_{\mp})+re^{\pm i\phi_3+i\delta}f(m^2_{\mp}, m^2_{\pm}), 
\end{equation}
where $f(m^2_+, m^2_-)$ is the amplitude of the \dkpp\ decay,
$r$ is the ratio of the magnitudes of the two interfering amplitudes, 
and $\delta$ is the strong phase difference between them. 
The \dkpp\ decay amplitude $f$ can be determined
from a large sample of flavor-tagged \dkpp\ decays 
produced in continuum $e^+e^-$ annihilation. Once $f$ is known, 
a simultaneous fit of $B^+$ and $B^-$ data allows the 
contributions of $r$, $\phi_3$ and $\delta$ to be separated. 
The method has a two-fold ambiguity: 
$(\phi_3,\delta)$ and $(\phi_3+180^{\circ}, \delta+180^{\circ})$
solutions cannot be separated. We always choose the solution 
with $0<\phi_3<180^{\circ}$. 
References \cite{giri} and \cite{belle_phi3_3} give  
a more detailed description of the technique. 

The method described above can be applied to other modes as well as \bdkp\ 
decay and its charge-conjugate mode (charge conjugate states are implied 
throughout the paper).
Excited states of neutral $D$ and $K$ mesons can also be used, although 
the values of $\delta$ and $r$ can differ for these decays. 
Both BaBar and Belle collaborations have successfully applied this 
technique to \bddsks\ modes with $D^0$ decaying to 
$K^0_S\pi^+\pi^-$~\cite{babar_phi3_2, babar_phi3_bdks, belle_phi3_1, belle_phi3_2, belle_phi3_3}. 
In addition, the BaBar collaboration reported a measurement of 
$\phi_3$ using the \bdk\ mode with $D^0$ decaying to the 
$\pi^0\pi^+\pi^-$ final state~\cite{babar_3pi}. 

In the current paper, we report a measurement of $\phi_3$ using the modes 
\bdkp\ and \bdskp\ with \dkpp, based on a 605 fb$^{-1}$ data sample
collected by the Belle detector at the KEKB asymmetric 
$e^+e^-$ factory. These results are preliminary. 

\section{Event selection}

We use a data sample of 
$657\times 10^6$ $B\overline{B}$ pairs, collected by the Belle detector. 
The decay chains \bdkp\ and \bdskp\ with $D^*\to D\pi^0$
are selected for the analysis. The neutral $D$ meson 
is reconstructed in the $K^0_S\pi^+\pi^-$ final state in all cases. 
We also select decays of \dsdpim\ produced via the 
$e^+e^-\to c\bar{c}$ continuum process as a high-statistics 
sample to determine the \dkpp\ decay amplitude. 

The Belle detector is described in detail elsewhere \cite{belle,svd2}. 
It is a large-solid-angle magnetic spectrometer consisting of a
silicon vertex detector (SVD), a 50-layer central drift chamber (CDC) for
charged particle tracking and specific ionization measurement ($dE/dx$), 
an array of aerogel threshold Cherenkov counters (ACC), time-of-flight
scintillation counters (TOF), and an array of CsI(Tl) crystals for 
electromagnetic calorimetry (ECL) located inside a superconducting solenoid coil
that provides a 1.5 T magnetic field. An iron flux return located outside 
the coil is instrumented to detect $K_L$ mesons and identify muons (KLM).

Charged tracks are required to satisfy criteria based on the 
quality of the track fit and the distance from the interaction point. 
We require each track to have a transverse momentum greater than 
100 MeV/$c$. Separation of kaons and pions is accomplished by combining 
the responses of 
the ACC and the TOF with the $dE/dx$ measurement from the CDC. 
Photon candidates are required to have ECL energy greater than 30 MeV. 
Neutral pion candidates are formed from pairs of photons with invariant 
masses in the range 120 to 150 MeV/$c^2$. 
Neutral kaons are reconstructed from pairs of oppositely charged tracks
with an invariant mass $M_{\pi\pi}$ within $7$ MeV/$c^2$ of the nominal 
$K^0_S$ mass and the reconstructed vertex distance from the 
interaction point greater than 1 mm. 

To determine the \dkpp\ decay amplitude we use $D^{*\pm}$ mesons
produced via the $e^+ e^-\to c\bar{c}$ continuum process. 
The flavor of the neutral $D$ meson is tagged by the charge of the slow pion 
(which we denote as $\pi_s$) in the decay \dsdpims.
The slow pion track is required to originate from the $D^0$ decay vertex to 
improve the momentum and angular resolution of the $\pi_s$. 
To select neutral $D$ candidates we require the invariant mass of the 
$K^0_S\pi^+\pi^-$ system to be within 11 MeV/$c^2$ of the $D^0$ mass.
To select events originating from a $D^{*-}$ decay 
we impose a requirement on the difference 
$\Delta M$ of the invariant 
masses of the $D^{*-}$ and the neutral $D$ candidates: 
$144.9\mbox{ MeV}/c^2<\Delta M<145.9\mbox{ MeV}/c^2$.
The suppression of the combinatorial background from $B\overline{B}$ events
is achieved by requiring the $D^{*-}$ momentum 
in the center-of-mass (CM) frame to be greater than 2.7 GeV/$c$.
The number of events in the signal region is $290.9\times 10^3$; 
the background fraction is 1.0\%.

The selection of $B$ candidates is based on the CM energy difference
$\Delta E = \sum E_i - E_{\rm beam}$ and the beam-constrained $B$ meson mass
$M_{\rm bc} = \sqrt{E_{\rm beam}^2 - (\sum \vec{p}_i)^2}$, where $E_{\rm beam}$ 
is the CM beam 
energy, and $E_i$ and $\vec{p}_i$ are the CM energies and momenta of the
$B$ candidate decay products. 
We also impose a requirement on the invariant mass of the neutral $D$ candidate: 
$|M_{K^0_S\pi^+\pi^-}-M_{D^0}|<11$~MeV/$c^2$. 

To suppress background from $e^+e^-\to q\bar{q}$ ($q=u, d, s, c$) 
continuum events, we calculate two variables which characterize the 
event shape. One is the cosine of the thrust angle $\cos\theta_{\rm thr}$, 
where $\theta_{\rm thr}$ is the angle between the thrust axis of 
the $B$ candidate daughters and that of the rest of the event. 
The other is a Fisher discriminant $\fish$ composed of 11 parameters \cite{fisher}: 
the production angle of the $B$ candidate, the angle of the $B$ thrust 
axis relative to the beam axis and nine parameters representing 
the momentum flow in the event relative to the $B$ thrust axis in the CM frame.
At the first stage of the analysis, when the $(\mbc, \de)$
distribution is fitted in order to obtain the fractions of the 
background components, we require $|\thr|<0.8$ and $\fish>-0.7$. 
In the Dalitz plot fit, we do not reject events based on these 
variables (as in the previous analysis~\cite{belle_phi3_3}), 
but rather use them in the likelihood function to 
better separate signal and background events. 
This leads to a 7--8\% improvement in 
the expected statistical error. 

\begin{figure}
  \epsfig{figure=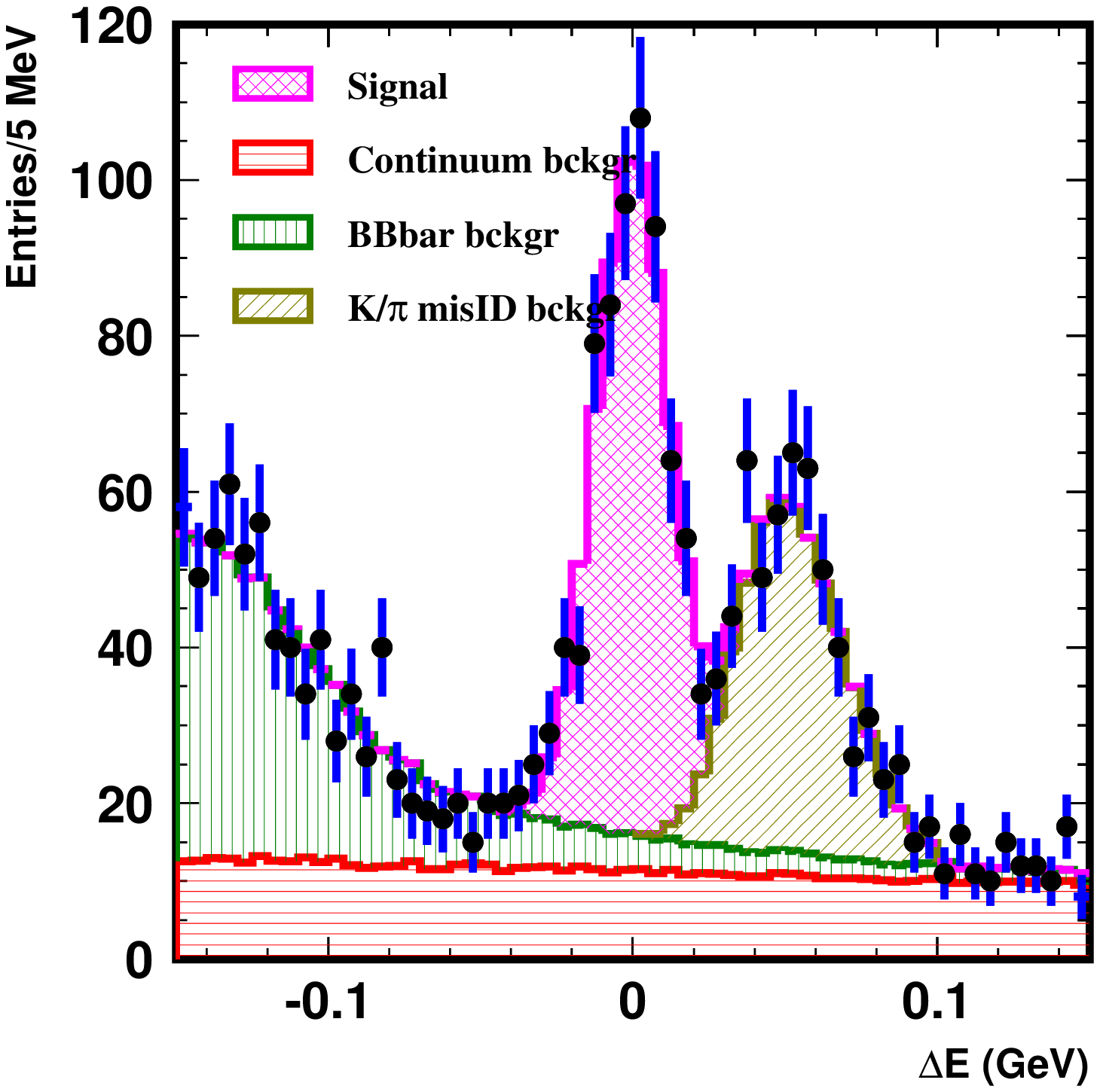,width=0.23\textwidth}
  \epsfig{figure=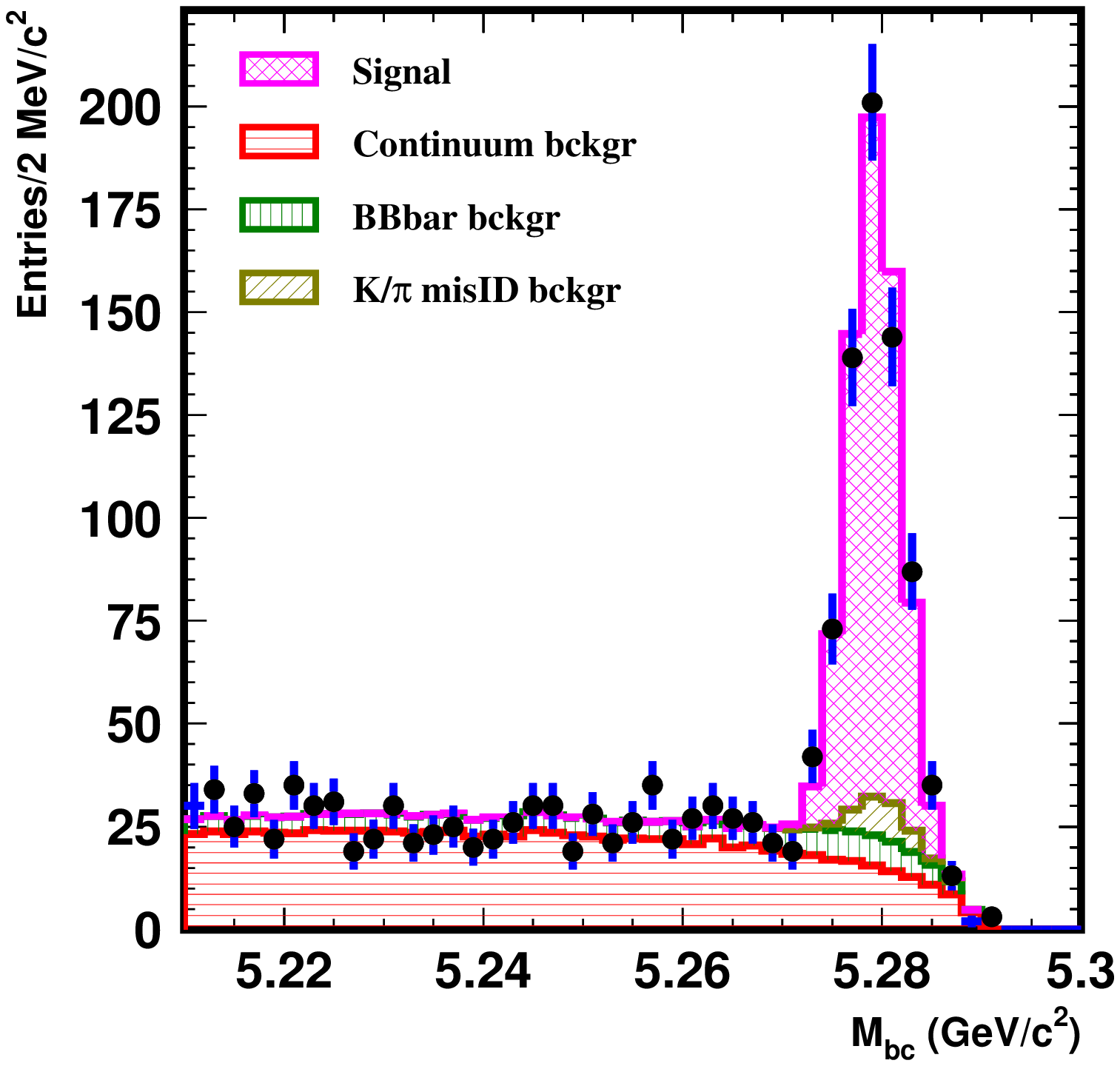,width=0.23\textwidth}
  \epsfig{figure=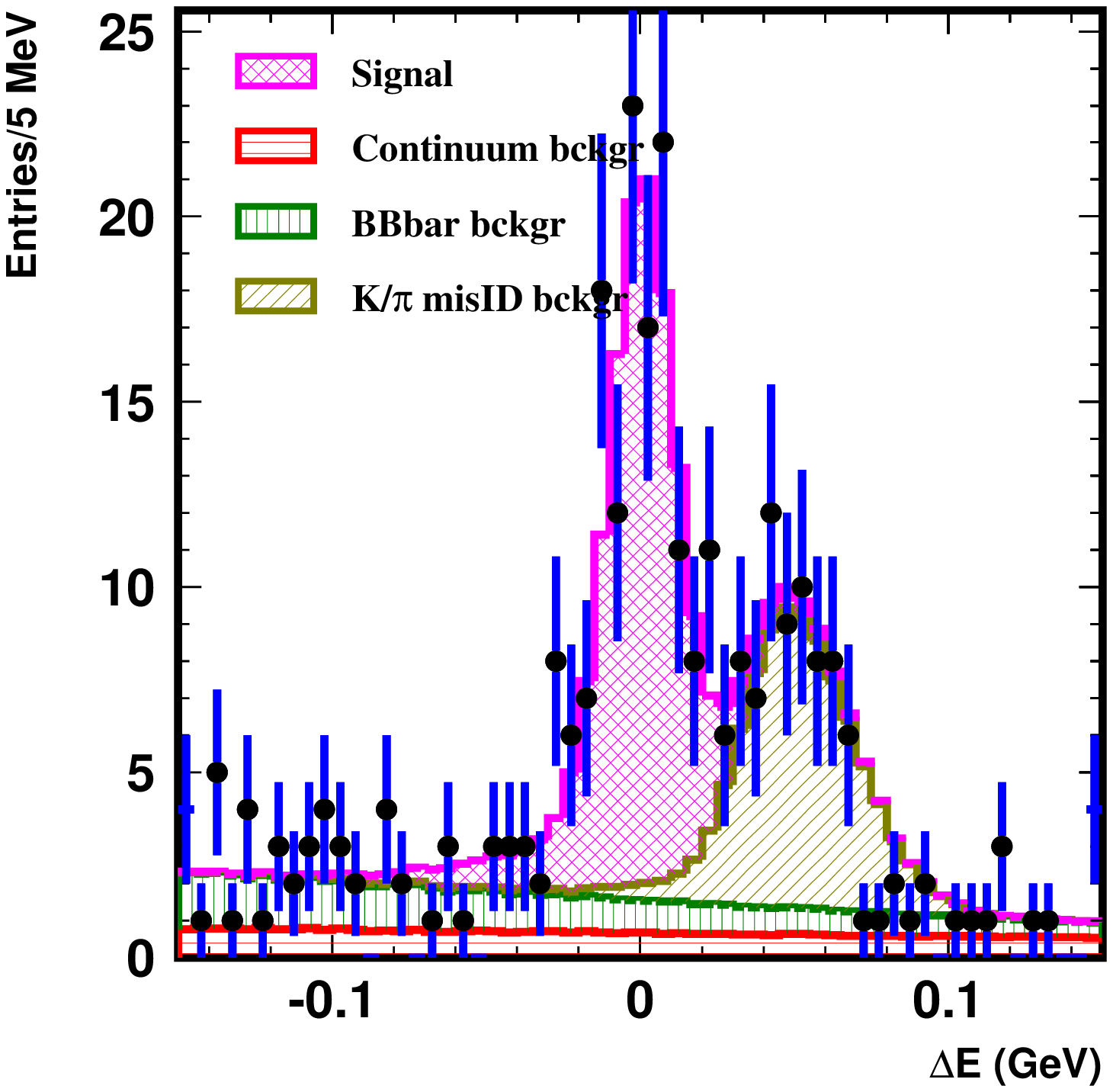,width=0.23\textwidth}
  \epsfig{figure=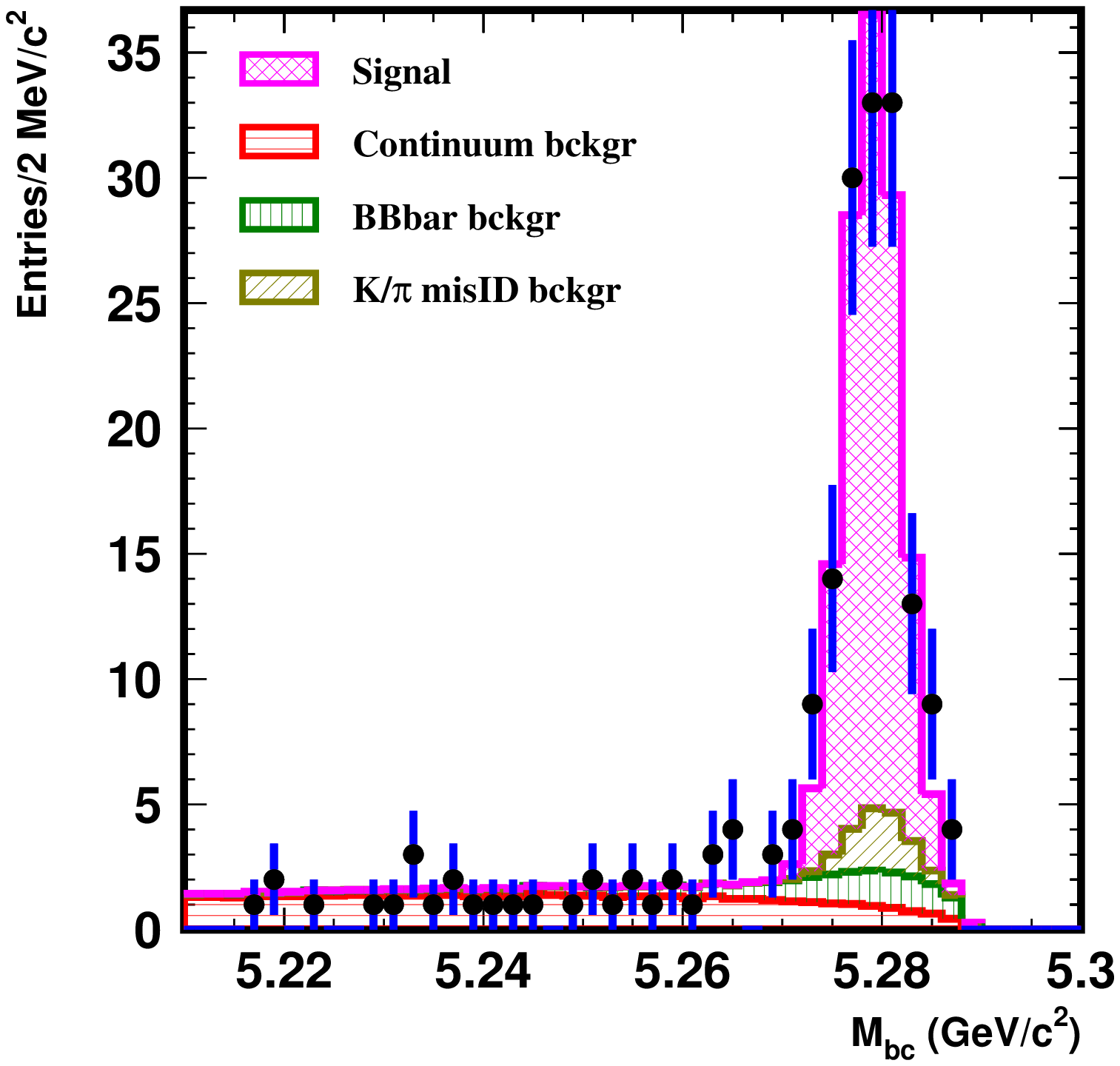,width=0.23\textwidth}
  \caption{$\Delta E$ and $M_{\rm bc}$ distributions for the 
           \bdkp\ (top) and \bdskp\ (bottom) event samples. 
           Points with error bars are the data, and
           the histogram is the result of a MC simulation according to the 
           fit result. The $\de$ ($\mbc$) distributions are shown here 
           with a signal-region selection of $\mbc>5.27$ GeV/$c^2$ 
           ($|\de|<30$ MeV) applied; this fit is performed on the full region.}
  \label{signal_mbcde}
\end{figure}

The $\Delta E$ and $M_{\rm bc}$ distributions for \bdkp\ and 
\bdskp\ candidates are shown in Fig.~\ref{signal_mbcde}. 
For the selected events a two-dimensional unbinned maximum 
likelihood fit in the variables $M_{\rm bc}$ and $\Delta E$ 
is performed, with the fractions of continuum, $B\bar{B}$ and \bddspi\ backgrounds
as free parameters, and their distributions fixed from generic MC 
simulation. The resulting signal and background density functions 
are used in the Dalitz plot fit to obtain the event-by-event signal 
to background ratio. The number of events in the signal box 
($\mbc>5.27$ GeV/$c^2$, $|\de|<30$ MeV, $|\thr|<0.8$, $\fish>-0.7$)
is 756. The $(\mbc,\de)$ fit yields a continuum background fraction of 
$(17.9\pm 0.7)$\%, $B\overline{B}$ background fraction of $(7.3\pm 0.5)$\%, 
and a \bdpi\ background fraction of $(4.3\pm 0.3)$\% in the signal box. 

To select \bdskp\ events, in addition to the 
requirements described above, we require that the mass difference 
$\Delta M$ of neutral $D^{*}$ and $D$ candidates satisfies 
$140\mbox{ MeV}/c^2<\Delta M<144\mbox{ MeV}/c^2$.
The number of events in the signal box is 149. 
The continuum background fraction is $(5.7\pm 0.7)$\%, the 
$B\overline{B}$ background fraction is $(7.6\pm 1.9)$\%, 
and \bdspi\ background fraction is $(7.0\pm 1.3)$\%. 

\begin{figure}
  \epsfig{figure=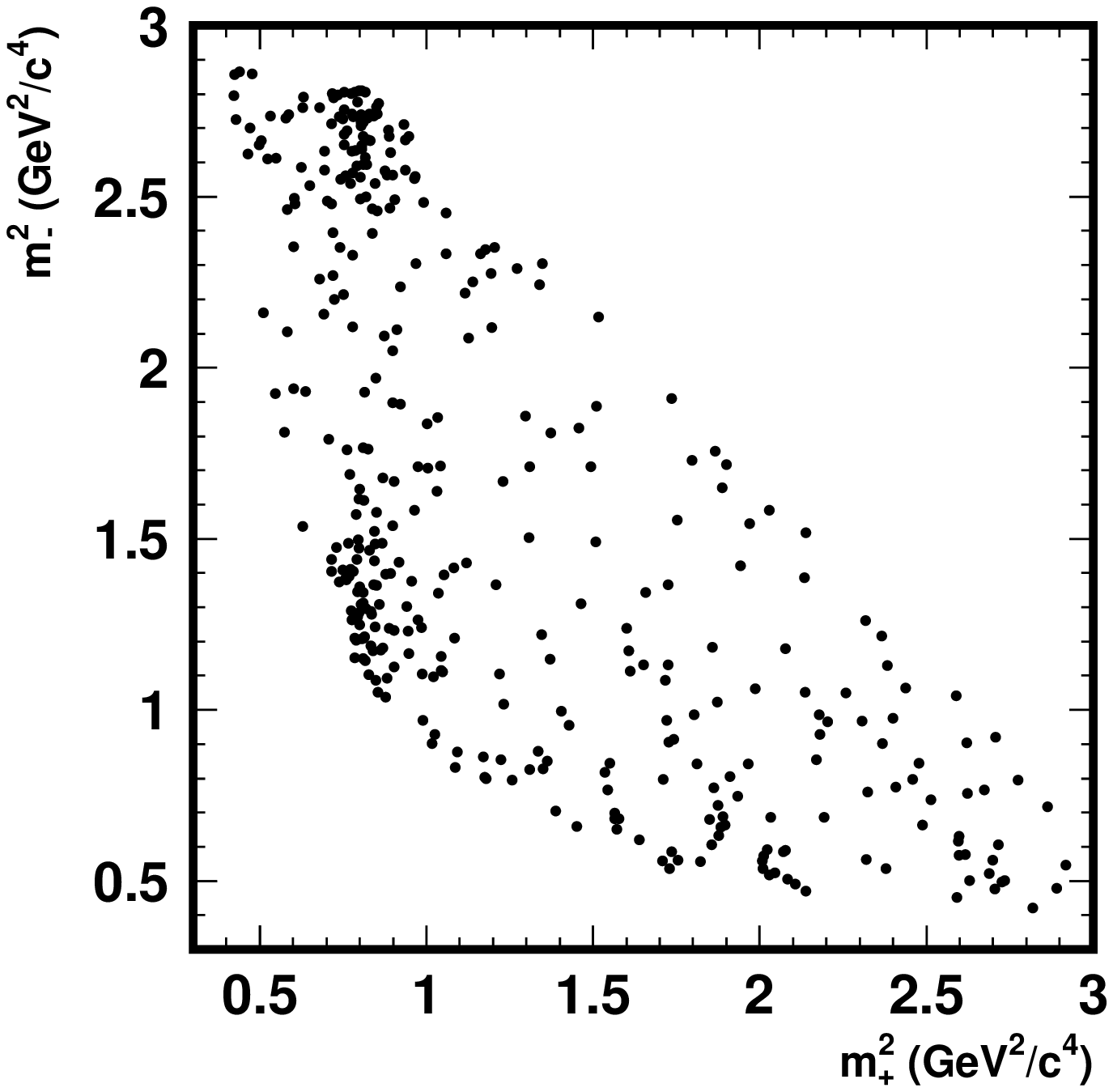,width=0.23\textwidth}
  \epsfig{figure=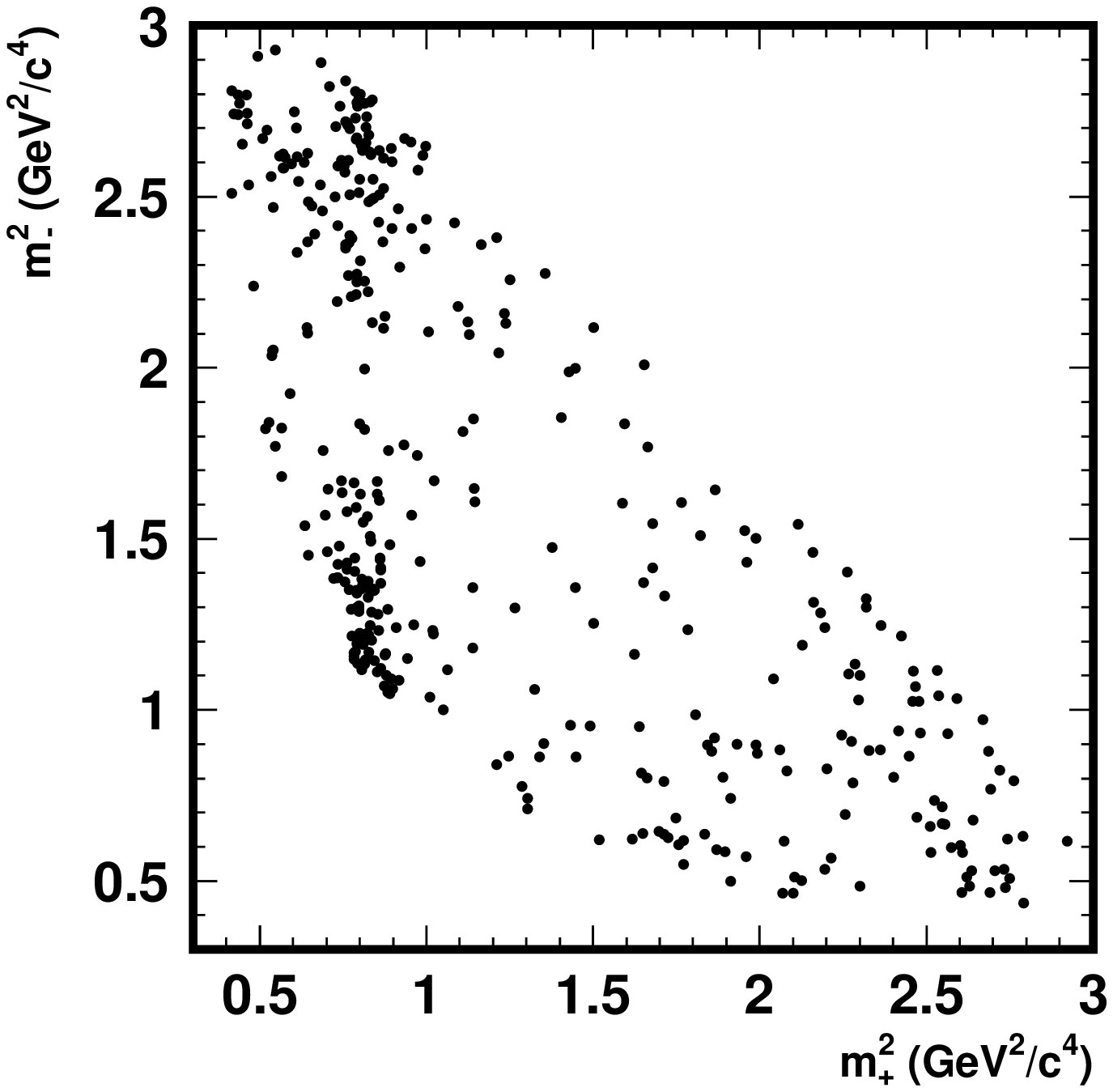,width=0.23\textwidth}
  \epsfig{figure=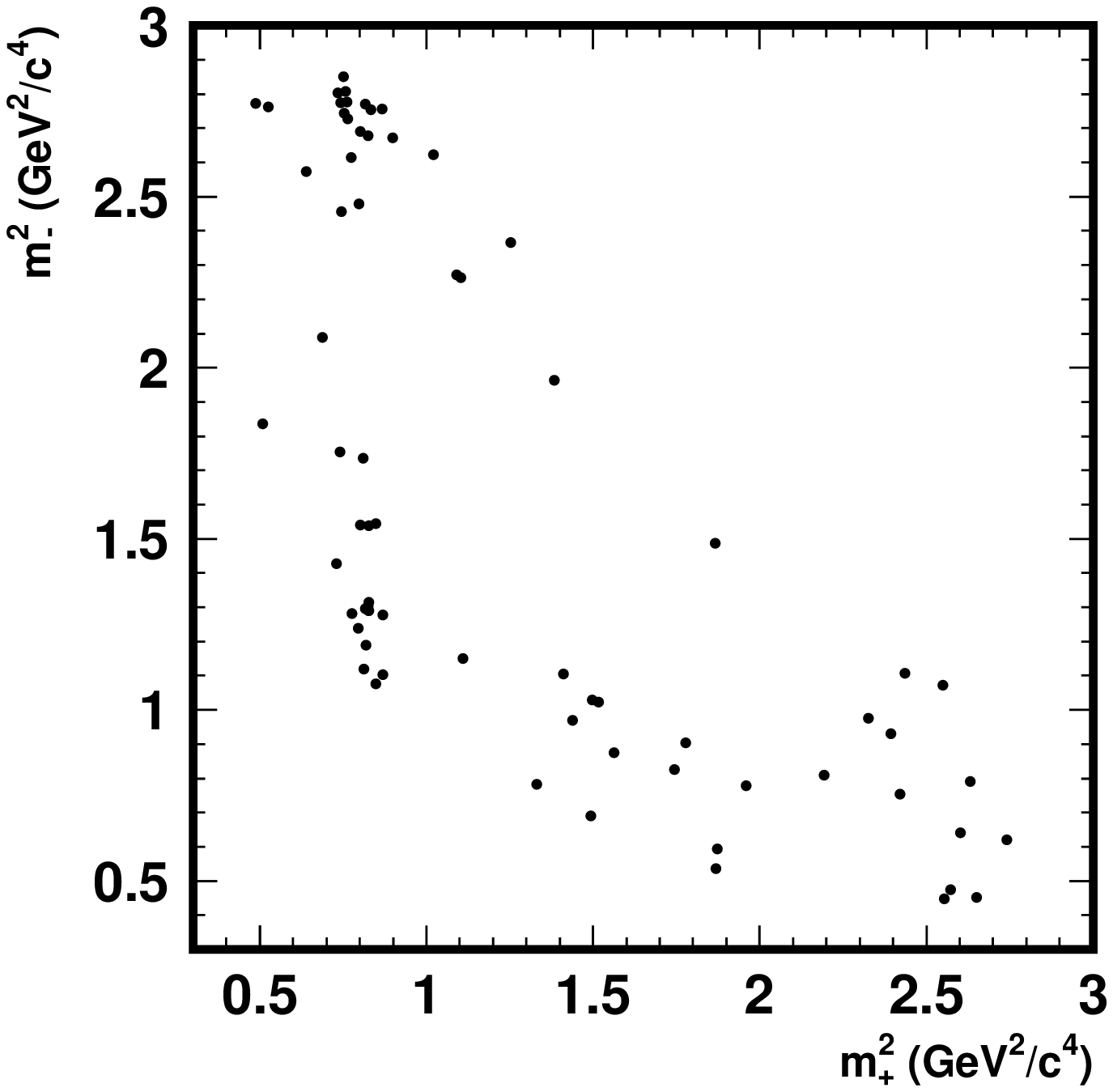,width=0.23\textwidth}
  \epsfig{figure=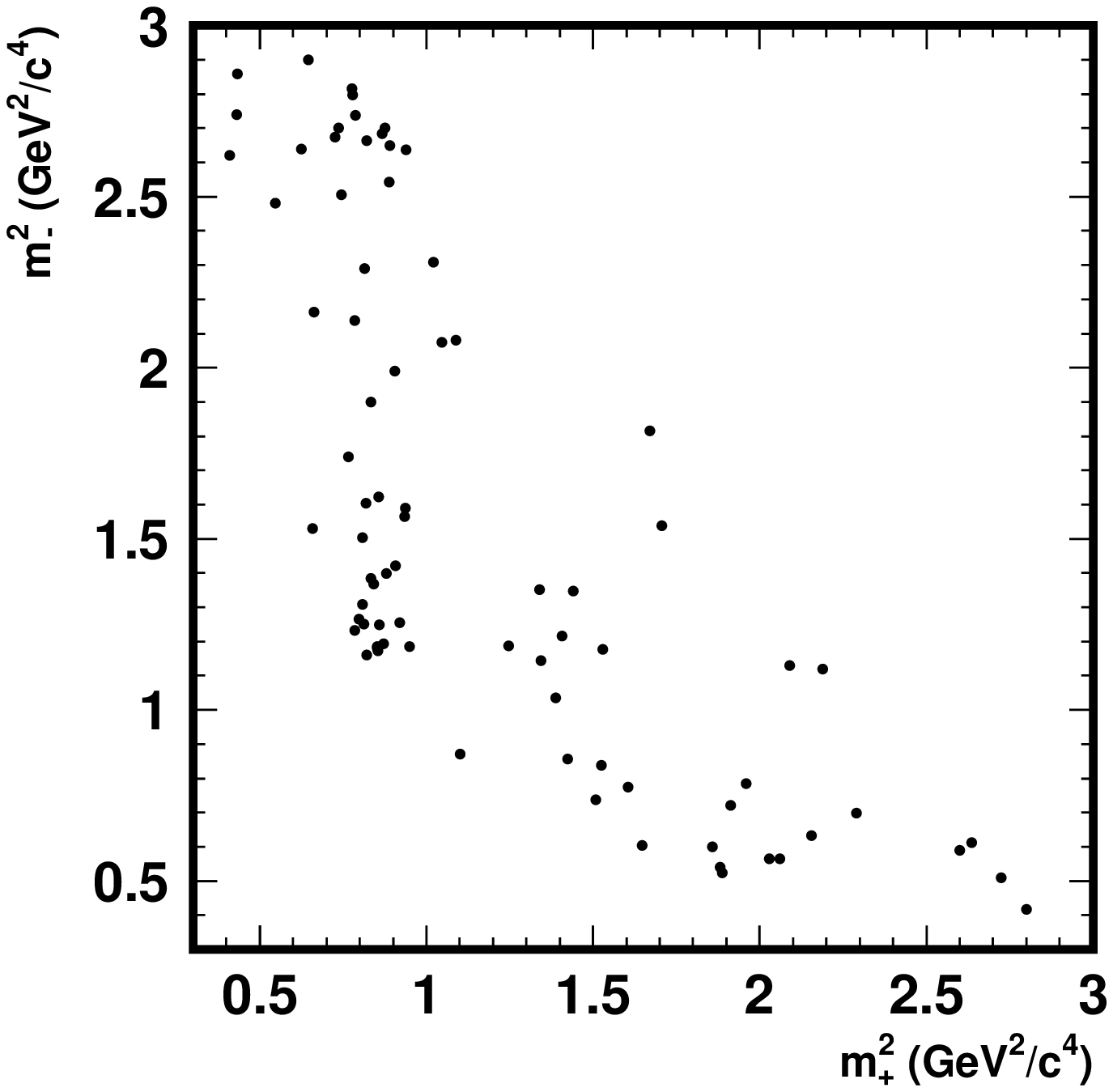,width=0.23\textwidth}
  \caption{Dalitz distributions of \dkpp\ decays from selected 
            \bdk\ (top) and \bdsk\ (bottom) candidates, 
            shown separately for $B^-$ (left) and $B^+$ (right) tags. }
  \label{dalitz_plots}
\end{figure}

The Dalitz distributions of \dkpp\ decay in the signal box of \bdk\ 
and \bdsk\ processes are shown in Fig.~\ref{dalitz_plots}. 

\section{Determination of the \boldmath{\dkpp} decay amplitude}

\label{section_d0_fit}

As in our previous analysis~\cite{belle_phi3_3}, 
the \dkpp\ decay amplitude is represented using the isobar model. 
The list of resonances 
is also the same, the only difference being the free parameters (mass and 
width) of the $K^*(892)^{\pm}$ and $\rho(770)$ states. 
A modified amplitude, where the scalar $\pi\pi$ component is described 
using the K-matrix approach~\cite{kmatrix}, is used in the estimation of 
the systematic error. 

The amplitude $f$ for the \dkpp\ decay is described
by a coherent sum of $N$ two-body decay amplitudes and one non-resonant 
decay amplitude,
\begin{equation}
  f(m^2_+, m^2_-) = \sum\limits_{j=1}^{N} a_j e^{i\xi_j}
  \mathcal{A}_j(m^2_+, m^2_-)+
    a_{\text{NR}} e^{i\xi_{\text{NR}}}, 
  \label{d0_model}
\end{equation}
where $\mathcal{A}_j(m^2_+, m^2_-)$ is the matrix element, $a_j$ and 
$\xi_j$ are the amplitude and phase of the matrix element, respectively, 
of the $j$-th resonance, and $a_{\text{NR}}$ and $\xi_{\text{NR}}$
are the amplitude and phase of the non-resonant component. 
The description of the matrix elements follows Ref.~\cite{cleo_model}. 
We use a set of 18 two-body amplitudes. 
These include five Cabibbo-allowed amplitudes: $K^*(892)^+\pi^-$, 
$K^*(1410)^+\pi^-$, $K_0^*(1430)^+\pi^-$, 
$K_2^*(1430)^+\pi^-$ and $K^*(1680)^+\pi^-$;  
their doubly Cabibbo-suppressed partners; and eight amplitudes with
$K^0_S$ and a $\pi\pi$ resonance:
$K^0_S\rho$, $K^0_S\omega$, $K^0_Sf_0(980)$, $K^0_Sf_2(1270)$, 
$K^0_Sf_0(1370)$, $K^0_S\rho(1450)$, $K^0_S\sigma_1$ and $K^0_S\sigma_2$. 

We use an unbinned maximum likelihood technique to fit the Dalitz plot 
distribution to the model described by Eq.~\ref{d0_model} with 
efficiency variation, background contributions and 
finite momentum resolution taken into account. 
The free parameters of the minimization are the amplitudes
$a_j$ and phases $\xi_j$ of the resonances, 
the amplitude $a_{NR}$ and phase $\xi_{NR}$ of the non-resonant component
and the masses and widths of the $\sigma_1$ and $\sigma_2$ scalars. 
We also allow the masses and widths of the $K^*(892)^+$ and $\rho(770)$
states to float. 

The procedures for determining the background
density, the efficiency, and the resolution 
are the same as in the previous analyses 
\cite{belle_phi3_2,belle_phi3_3}. 
The background density for \dkpp\ events is extracted from 
$\Delta M$ sidebands. The shape of the efficiency over the Dalitz plot, 
as well as the invariant mass resolution, is extracted from the 
signal Monte-Carlo (MC) simulation. 

The fit results are given in Table~\ref{dkpp_table}. 
The parameters obtained for the $\sigma_1$ resonance 
($M_{\sigma_1}=522\pm 6$ MeV/$c^2$, $\Gamma_{\sigma_1}=453\pm 10$ MeV/$c^2$) 
are similar to those observed by other experiments~\cite{dkpp_cleo, aitala2}.
The second scalar term $\sigma_2$ is introduced to account for
structure observed at $m^2_{\pi\pi} \sim 1.1\,\mathrm{GeV}^2/c^4$:
the fit finds a small but significant contribution with
$M_{\sigma_2}=1033\pm 7$ MeV/$c^2$, $\Gamma_{\sigma_2}=88\pm 7$ MeV/$c^2$.
Allowing the parameters of the dominant $K^*(892)^+$ and $\rho(770)$
resonances to float results in a significant improvement in the fit quality. 
We obtain $M(K^*(892))=893.7\pm 0.1$ MeV/$c^2$, 
$\Gamma(K^*(892))=48.4\pm 0.2$ MeV/$c^2$, 
$M(\rho)=771.7\pm 0.7$ MeV/$c^2$, 
$\Gamma(\rho)=136.0\pm 1.3$ MeV/$c^2$. 

The $\chi^2$ test finds a value of 
$\chi^2/ndf=2.35$ for 1065 degrees of freedom ($ndf$), which is large. 
We find that the main features of the 
Dalitz plot are well-reproduced, with some significant but numerically
small discrepancies at peaks and dips of the distribution. 
In our final results we include a conservative contribution to the 
systematic error due to uncertainties in the $\overline{D}{}^0$ decay model. 
 
\begin{table}
\caption{Fit results for \dkpp\ decay. Errors are statistical only.}
\label{dkpp_table}
\begin{tabular}{|l|c|c|} \hline
Intermediate state           & Amplitude 
			     & Phase ($^{\circ}$) 
			     \\ \hline

$K_S \sigma_1$               & $1.56\pm 0.06$
                             & $214\pm 3$
                             \\

$K_S\rho^0$                  & $1.0$ (fixed)                                 
                             & 0 (fixed)   
                             \\

$K_S\omega$                  & $0.0343\pm 0.0008$
                             & $112.0\pm 1.3$
                             \\

$K_S f_0(980)$               & $0.385\pm 0.006$
                             & $207.3\pm 2.3$
                             \\

$K_S \sigma_2$               & $0.20\pm 0.02$
                             & $212\pm 12$
                             \\

$K_S f_2(1270)$              & $1.44\pm 0.04$
                             & $342.9\pm 1.7$
                             \\

$K_S f_0(1370)$              & $1.56\pm 0.12$ 
                             & $110\pm 4$
                             \\

$K_S \rho^0(1450)$           & $0.49\pm 0.08$
                             & $64\pm 11$
                             \\

$K^*(892)^+\pi^-$            & $1.638\pm 0.010$
                             & $133.2\pm 0.4$
                             \\ 

$K^*(892)^-\pi^+$            & $0.149\pm 0.004$
                             & $325.4\pm 1.3$
                             \\

$K^*(1410)^+\pi^-$	     & $0.65\pm 0.05$
			     & $120\pm 4$
			     \\

$K^*(1410)^-\pi^+$	     & $0.42\pm 0.04$
			     & $253\pm 5$
			     \\

$K_0^*(1430)^+\pi^-$         & $2.21\pm 0.04$
                             & $358.9\pm 1.1$
                             \\

$K_0^*(1430)^-\pi^+$         & $0.36\pm 0.03$
                             & $87\pm 4$
                             \\

$K_2^*(1430)^+\pi^-$         & $0.89\pm 0.03$
                             & $314.8\pm 1.1$
                             \\

$K_2^*(1430)^-\pi^+$         & $0.23\pm 0.02$
                             & $275\pm 6$
                             \\

$K^*(1680)^+\pi^-$           & $0.88\pm 0.27$
                             & $82\pm 17$
                             \\

$K^*(1680)^-\pi^+$           & $2.1\pm 0.2$
                             & $130\pm 6$
                             \\

non-resonant                 & $2.7\pm 0.3$
                             & $160\pm 5$
                             \\ 
\hline
\end{tabular}
\end{table}

\section{Dalitz plot analysis of \boldmath{\bddskp} decays}

\label{dalitz_analysis}

As in our previous analysis~\cite{belle_phi3_3} and in analyses 
carried out by the BaBar collaboration~\cite{babar_phi3_2, babar_phi3_bdks}, 
we fit the Dalitz distributions of the 
$B^+$ and $B^-$ samples separately, using Cartesian parameters 
$x_{\pm}=r_{\pm}\cos(\pm\phi_3+\delta)$ and 
$y_{\pm}=r_{\pm}\sin(\pm\phi_3+\delta)$, where the indices ``$+$" and 
``$-$" correspond to $B^+$ and $B^-$ decays, respectively. 
In this approach the amplitude ratios ($r_+$ and $r_-$) are 
not constrained to be equal for the $B^+$ and $B^-$ samples. 
Confidence intervals in $r$, $\phi_3$ and $\delta$ are then obtained 
from the $(x_{\pm},y_{\pm})$ using a frequentist technique. The advantage
of this approach is low bias and simple distributions of the fitted 
parameters, at the price of fitting in a space with higher dimensionality 
$(x_+,y_+,x_-,y_-)$ than that of the physical parameters 
$(r, \phi_3, \delta)$; see Section~\ref{section_stat}. 

The fit to a single Dalitz distribution with free parameters 
$x$ and $y$ is performed by the unbinned maximum likelihood technique, 
using variables $m^2_+$, $m^2_-$, $M_{\rm bc}$, $\Delta E$, $\thr$ and 
$\fish$; 
only the first four variables 
were used in the previous analysis~\cite{belle_phi3_3}. 
We subdivide the background distribution into four components:
$e^+e^-\to q\bar{q}$ (where $q=u,d,s$), 
charm, $B\overline{B}$ (except for \bddspi) 
and \bddspi\ background. 
The distributions of each of these components are assumed to be 
factorized as products of a Dalitz plot distribution $(m^2_+, m_-^2)$, 
and distributions in $(\mbc, \de)$, and $(\thr,\fish)$. 
The shapes of these distributions are extracted from MC simulation. 
The six-dimensional PDF used for the fit is thus expressed as
\begin{equation}
  p=\sum\limits_i p_i(m_+^2, m_-^2)p_i(\mbc, \de)p_i(\thr, \fish), 
  \label{prob_dens}
\end{equation}
where the index $i$ runs over all background contributions and signal. 
Possible deviations from the factorization assumption and disagreements 
between MC and experimental background densities are treated in the 
systematic error. The efficiency variation as a function of the Dalitz 
plot variables is obtained from signal MC simulation and is taken into 
account in the likelihood function. 

To test the consistency of the fit, the same procedure as 
used for \bddskp\ signal was applied to the \bddspip\ control 
samples. The results are consistent with the expected value $r\sim 0.01$
for the amplitude ratio. 

\begin{table}
  \caption{Results of the signal fits in parameters $(x,y)$. The first error 
  is statistical, the second is experimental systematic error. 
  Model uncertainty is not included. }
  \label{sig_fit_table}
  \begin{tabular}{|l|c|c|}
  \hline
  Parameter  & \bdkp & \bdskp \\ 
  \hline
  $x_-$ & $+0.105\pm 0.047\pm 0.011$ & $+0.024\pm 0.140\pm 0.018$ \\
  $y_-$ & $+0.177\pm 0.060\pm 0.018$ & $-0.243\pm 0.137\pm 0.022$ \\
  $x_+$ & $-0.107\pm 0.043\pm 0.011$ & $+0.133\pm 0.083\pm 0.018$ \\
  $y_+$ & $-0.067\pm 0.059\pm 0.018$ & $+0.130\pm 0.120\pm 0.022$ \\ 
  \hline
  \end{tabular}
\end{table}

\begin{figure}
  \epsfig{figure=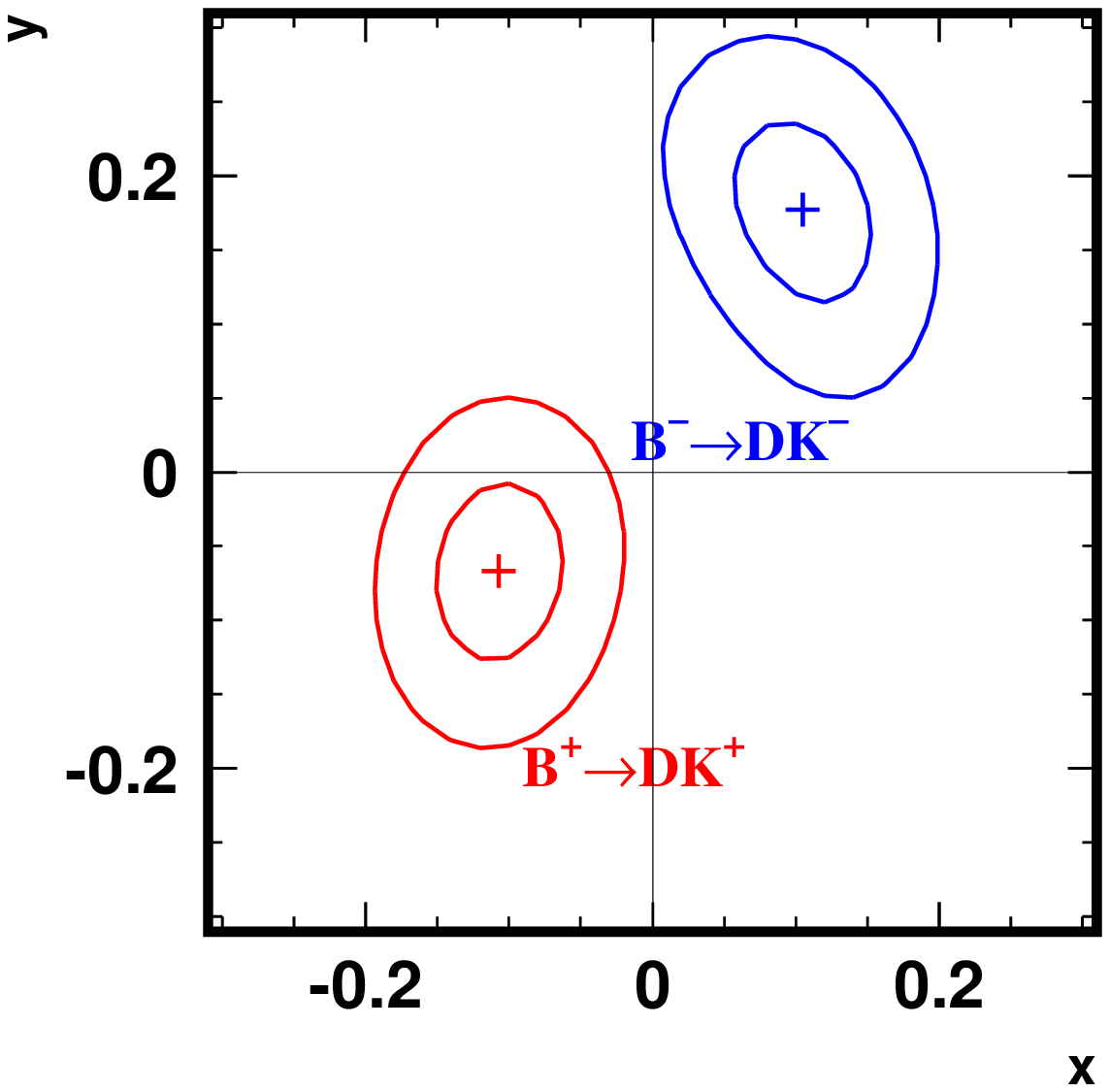,width=0.23\textwidth}
  \epsfig{figure=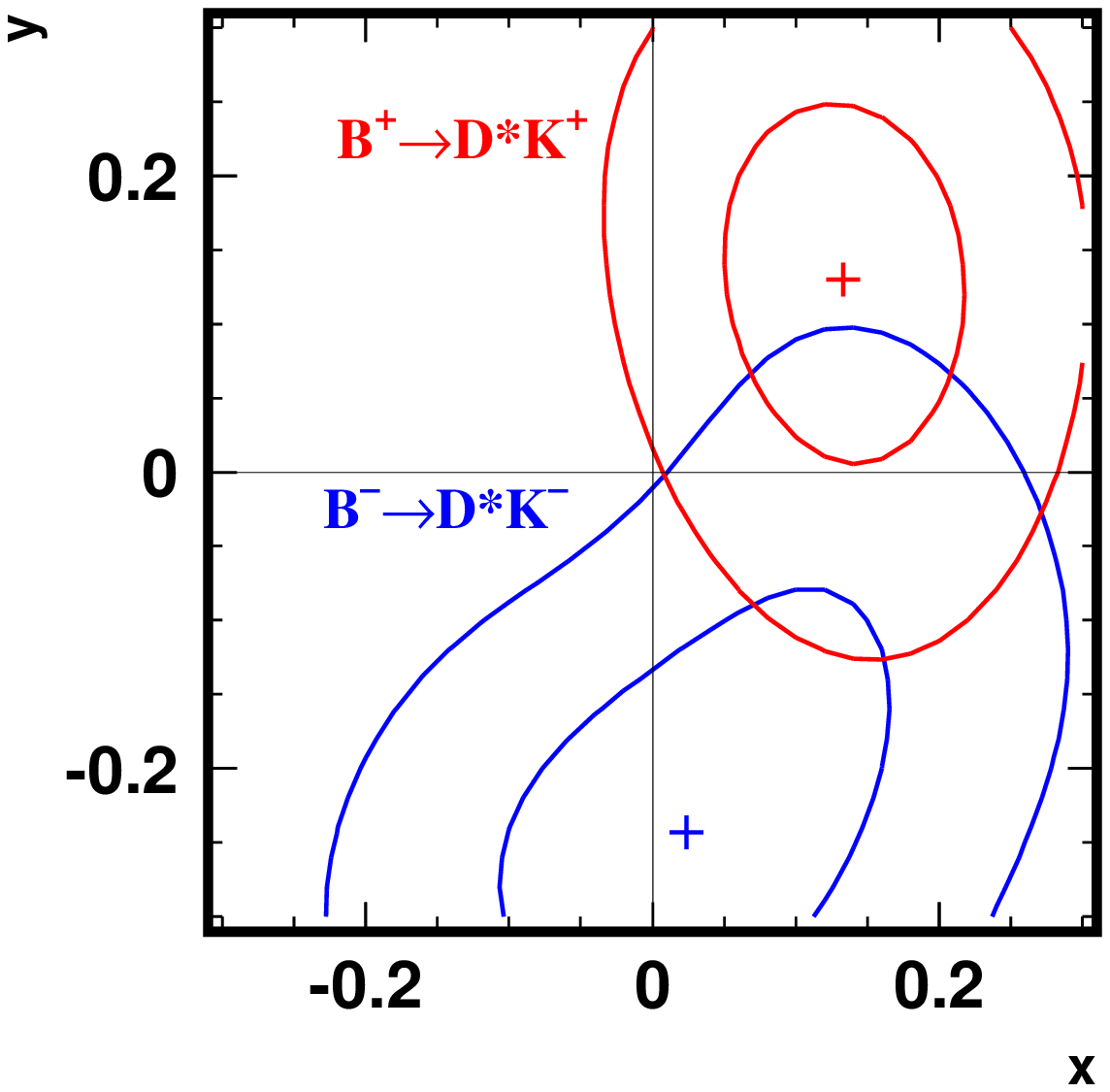,width=0.23\textwidth}
  \caption{Results of signal fits with free parameters 
           $x=r\cos\theta$ and $y=r\sin\theta$ for \bdkp\ (left), and 
           \bdskp\ (right) samples, separately for $B^-$
           and $B^+$ data. Contours indicate one and two
           standard deviation regions obtained in the maximum 
           likelihood fit. }
  \label{sig_fit}
\end{figure}

The results of the separate $B^+$ and $B^-$ data fits are shown in 
Fig.~\ref{sig_fit}. 
The values of the fit parameters $x_{\pm}$ and $y_{\pm}$ are 
listed in Table~\ref{sig_fit_table}. 

\section{Evaluation of the statistical errors}

\label{section_stat}

We use a frequentist technique to evaluate the 
statistical significance of the measurements. The procedure is identical 
to that in our previous analysis~\cite{belle_phi3_3}. 
This method requires knowledge of the probability density function (PDF) of the 
reconstructed parameters $x$ and $y$ as a function of the true parameters
$\bar{x}$ and $\bar{y}$. 
To obtain this PDF, we employ a ``toy" MC technique that uses a
simplified MC simulation of the experiment which incorporates
the maximum likelihood fit with the same efficiencies, resolution 
and backgrounds as used in the fit to the experimental data. 

\begin{table*}
  \caption{$CP$ fit results. The first error is statistical, the second 
  is experimental systematic, and the third is the model uncertainty. }
  \label{fit_res_table}
  \begin{tabular}{|l||c|c|} \hline
  Parameter & \bdkp\ mode
            & \bdskp\ mode\\ \hline
  $\phi_3$  & $80.8^{\circ}\;^{+13.1^{\circ}}_{-14.8^{\circ}}\pm 5.0^{\circ}\pm 8.7^{\circ}$ 
            & $63.8^{\circ}\;^{+20.8^{\circ}}_{-22.9^{\circ}}\pm 4.7^{\circ}\pm 8.7^{\circ}$ 
            \\
  $r$       & $0.161^{+0.040}_{-0.038}\pm 0.011\pm 0.049$
            & $0.208^{+0.085}_{-0.083}\pm 0.015\pm 0.049$
            \\
  $\delta$  & $137.4^{\circ}\;^{+13.0^{\circ}}_{-15.7^{\circ}}\pm 4.0^{\circ}\pm 22.9^{\circ}$
            & $342.0^{\circ}\;^{+21.4^{\circ}}_{-22.9^{\circ}}\pm 3.7^{\circ}\pm 22.9^{\circ}$
            \\
  \hline
  \end{tabular}
\end{table*}

\begin{figure}
  \epsfig{figure=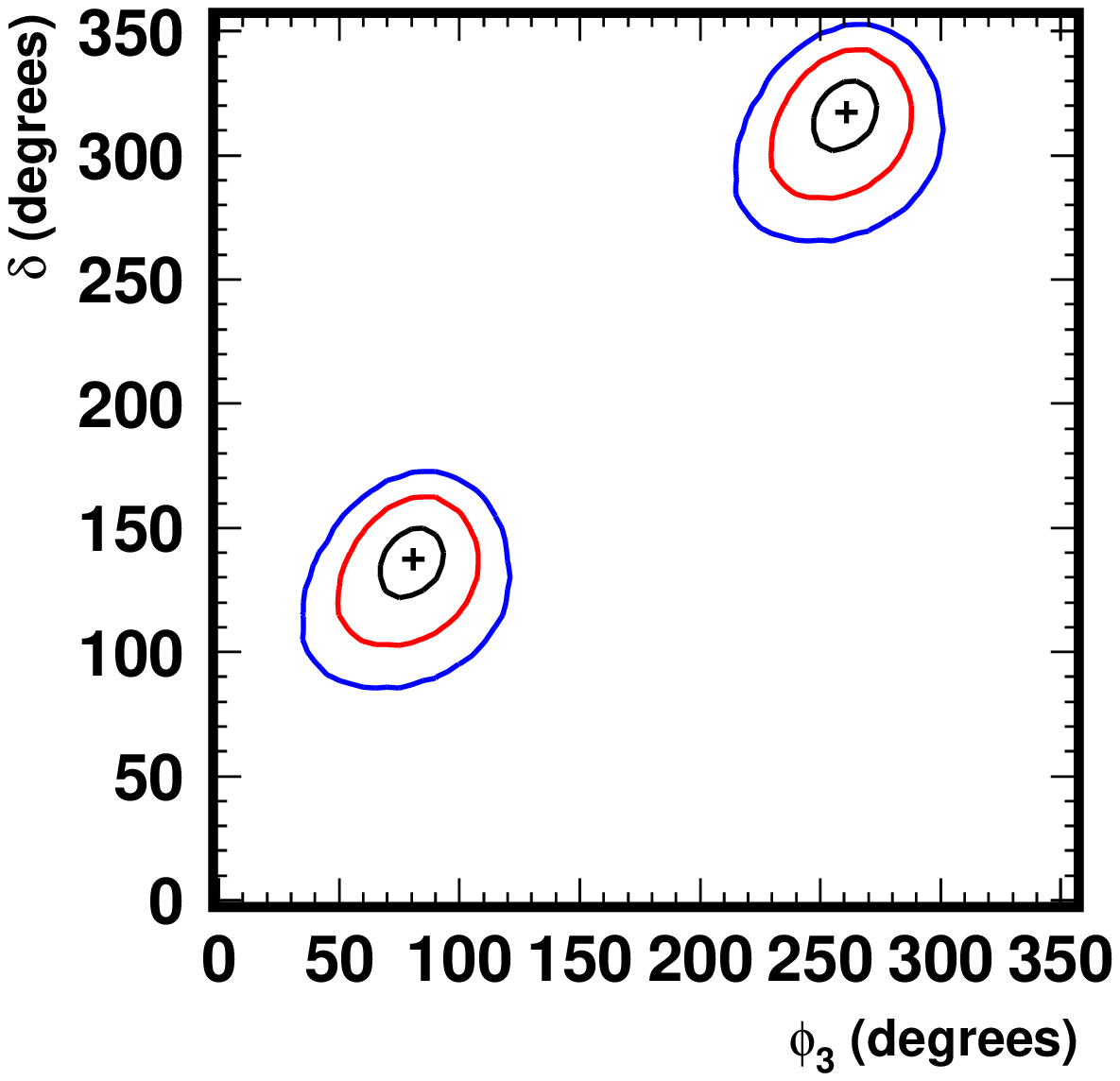,width=0.23\textwidth}
  \epsfig{figure=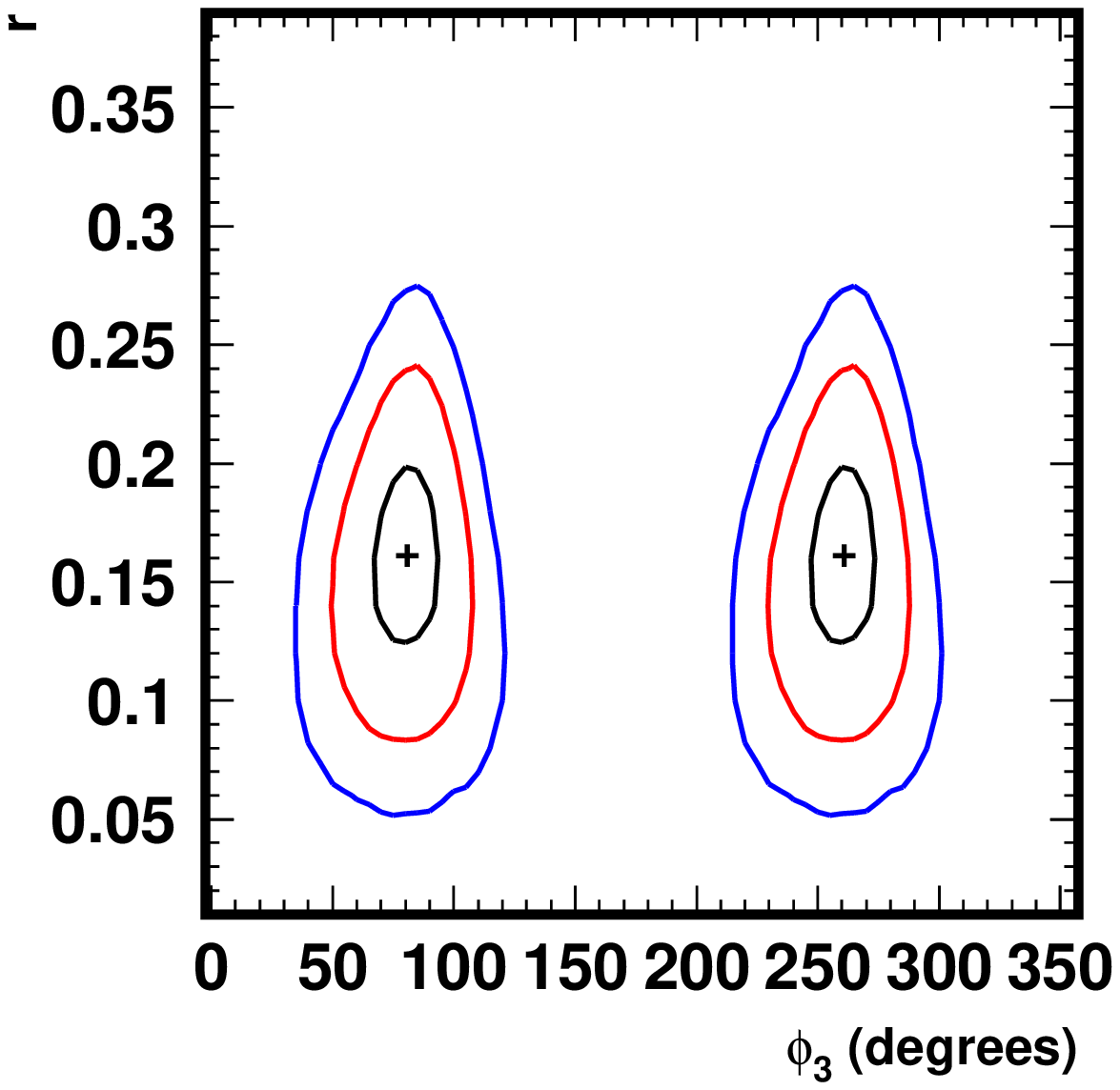,width=0.23\textwidth}
  \epsfig{figure=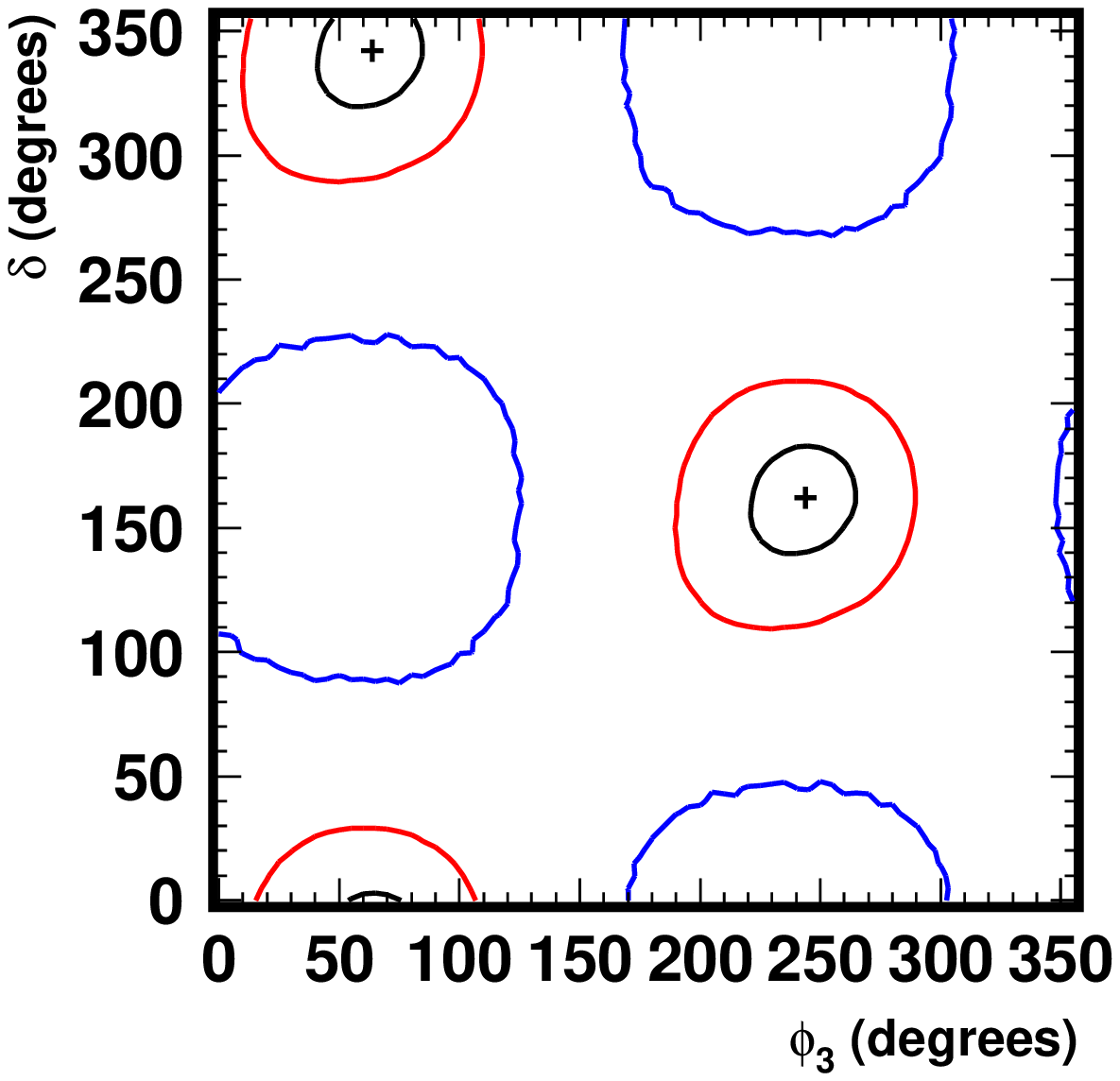,width=0.23\textwidth}
  \epsfig{figure=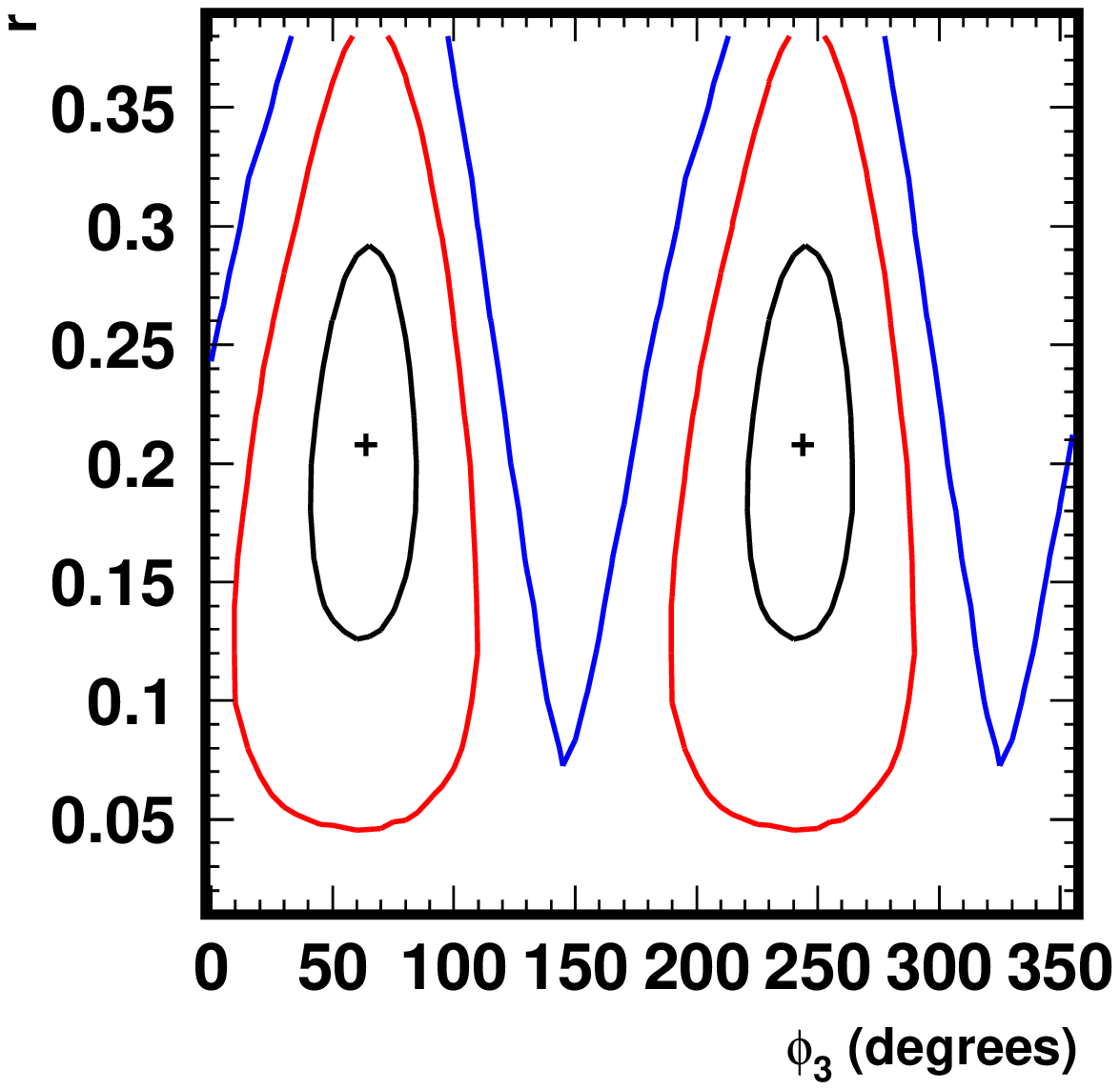,width=0.23\textwidth}
  \caption{Projections of confidence regions for the \bdkp\ (top) and \bdskp\ (bottom) 
           mode onto the $(r, \phi_3)$ and $(\phi_3, \delta)$ planes. 
           Contours indicate projections of one, two and three standard 
           deviation regions. }
  \label{cont_fc}
\end{figure}

Figure~\ref{cont_fc} shows the projections of the three-dimensional 
confidence regions onto the $(r, \phi_3)$ and $(\phi_3, \delta)$ planes
for \bdk\ and \bdsk\ modes. 
We show the 20\%, 74\% and 97\% confidence level regions, 
which correspond to one, two, and three standard deviations for a 
three-dimensional Gaussian distribution. The values of the 
parameters $r$, $\phi_3$ and $\delta$ obtained for \bdk\ and 
\bdsk\ modes separately are presented in Table~\ref{fit_res_table}. 

\section{Estimation of systematic error}

\label{section_syst}

Experimental systematic errors come from the uncertainty in the knowledge 
of the distributions used in the fit ({\it i.e.} Dalitz plot distributions 
of the background components, and the $(\mbc,\de)$ and 
$(\thr,\fish)$ distributions 
of the backgrounds and signal), fractions of different background components, 
and the distribution of the efficiency across the Dalitz plot phase space. 
Uncertainties due to background shapes are estimated by using alternative 
distributions in the fit (extracted from experimental data where possible). 
Uncertainties due to the background fractions are obtained by varying each 
fraction within its error. 
The procedure for estimating the uncertainty due to 
the detection efficiency is modified compared to the previous 
analysis~\cite{belle_phi3_3}: we use an 
alternative efficiency shape obtained by toy MC from the parameterized 
track finding efficiency (obtained from experimental data) as a function 
of transverse momentum and azimuthal angle $\theta$. 

Compared to our previous analysis~\cite{belle_phi3_3}, an additional 
source of systematic error exists due to the use of $\thr$ and $\fish$
variables in the fit. However, the use of these variables increases the 
effective signal-to-background ratio, so the total systematic error
is comparable. 

Systematic errors in the physical parameters $r$, $\phi_3$
and $\delta$ are calculated from the systematic errors on the 
fitted parameters $(x,y)$. Values $(x,y)$ are generated according to 
Gaussian distributions with standard deviations equal to the 
corresponding total systematic errors, then parameters $r$, $\phi_3$
and $\delta$ are obtained for each $(x,y)$ set, and the root-mean-square 
deviations (RMS) of the resulting values are calculated. 
We perform this procedure in two ways: without correlation of 
$(x,y)$ biases for $B^+$ and $B^-$, and with 100\% correlation between them. 
The largest RMS of the two options serves as the systematic error. 
The systematic errors in $x,y$ variables are shown in 
Table~\ref{sig_fit_table}. 

The model used for the \dkpp\ decay amplitude is one of the main sources of 
error for our analysis: we list this contribution separately. 
For the current measurement we use 
the same estimates of the model uncertainty as in our previous 
analysis~\cite{belle_phi3_3}. In addition, we perform a fit of the 
\dkpp\ amplitude using the K-matrix formalism~\cite{kmatrix} to describe 
the $\pi\pi$ $S$-wave contribution. The maximum difference between the 
baseline 
quasi two-body amplitude and the K-matrix amplitude in the $\phi_3$ fit 
is $\sim 2^{\circ}$. However, since the K-matrix describes only part 
of the amplitude, we still use $\Delta\phi_3=9^{\circ}$ as the estimate 
of the model uncertainty. 

\section{Combined \boldmath{$\phi_3$} measurement}

\label{section_combined}

The two event samples, \bdkp\ and \bdskp\, are combined 
in order to improve the sensitivity to $\phi_3$. 
The confidence levels for the combination of two modes are obtained 
using the frequentist technique as for the 
single mode, with the PDF of the two measurements being the product 
of the probability densities for the individual modes. 
Confidence intervals for the combined measurement together 
with systematic and model errors are shown in Table~\ref{fc_comb_table}. 
The statistical confidence level of $CP$ violation 
is $(1-5.5\times 10^{-4})$, or 3.5 standard deviations.

\begin{table*}
  \caption{Results of the combination of \bdkp\ and \bdskp\ modes. }
  \label{fc_comb_table}
  \begin{tabular}{|l|c|c|c|c|c|} \hline
  Parameter & $1\sigma$ interval & $2\sigma$ interval & 
              Systematic error & Model uncertainty \\ \hline
  $\phi_3$  & $76^{\circ}\;^{+12^{\circ}}_{-13^{\circ}}$ 
            & $49^{\circ}<\phi_3<99^{\circ}$ 
            & $4^{\circ}$ & $9^{\circ}$ \\
  $r_{DK}$  & $0.16\pm 0.04$ 
            & $0.08<r_{DK}<0.24$ 
            & $0.01$ & $0.05$ \\
  $r_{D^*K}$  & $0.21\pm 0.08$ 
            & $0.05<r_{D^*K}<0.39$ 
            & $0.02$ & $0.05$ \\
  $\delta_{DK}$  & $136^{\circ}\;^{+14^{\circ}}_{-16^{\circ}}$ 
            & $100^{\circ}<\delta_{DK}<163^{\circ}$ 
            & $4^{\circ}$ & $23^{\circ}$ \\
  $\delta_{D^*K}$  & $343^{\circ}\;^{+20^{\circ}}_{-22^{\circ}}$ 
            & $293^{\circ}<\delta_{DK}<389^{\circ}$ 
            & $4^{\circ}$ & $23^{\circ}$ \\
  \hline
  \end{tabular}
\end{table*}

\section{Conclusion}

We report the results of a measurement of the unitarity triangle angle 
$\phi_3$, using a method based on Dalitz plot analysis of
\dkpp\ decay in the process \bddskp. An updated
measurement of 
$\phi_3$ using this technique was performed based on 605 fb$^{-1}$ 
of data collected by the Belle detector: 70\% larger than the previous 
sample. The statistical sensitivity of the measurement has also been 
improved by modifications to the event selection and fit procedure. 


From the combination of \bdkp\ and \bdskp\ modes, we obtain the value
$\phi_3=76^{\circ}\;^{+12^{\circ}}_{-13^{\circ}}
\mbox{(stat)}\pm 4^{\circ} \mbox{(syst)}\pm 9^{\circ}(\mbox{model})$; 
of the two possible solutions we choose the one with $0<\phi_3<180^{\circ}$.
We also obtain values of the amplitude ratios 
$r_{DK}=0.16\pm 0.04\mbox{(stat)}\pm 0.01\mbox{(syst)}\pm 0.05\mbox{(model)}$
and $r_{D^*K}=0.21\pm 0.08\mbox{(stat)}\pm 0.02\mbox{(syst)}\pm 0.05\mbox{(model)}$. 
The statistical significance of $CP$ violation for the combined 
measurement is $(1-5.5\times 10^{-4})$, or 3.5 standard deviations. 
These results are preliminary. 

The statistical precision of the $\phi_3$ measurement is already
comparable to the estimated model uncertainty. However, it is possible
to eliminate this uncertainty using decays of $\psi(3770)\to
D^0\overline{D}{}^0$ \cite{modind,modind2}.

\section*{Acknowledgments}

We thank the KEKB group for excellent operation of the
accelerator, the KEK cryogenics group for efficient solenoid
operations, and the KEK computer group and
the NII for valuable computing and SINET3 network
support.  We acknowledge support from MEXT and JSPS (Japan);
ARC and DEST (Australia); NSFC and KIP of CAS (China); 
DST (India); MOEHRD, KOSEF and KRF (Korea); 
KBN (Poland); MES and RFAAE (Russia); ARRS (Slovenia); SNSF (Switzerland); 
NSC and MOE (Taiwan); and DOE (USA).

\end{document}